\documentclass[10pt]{article}
\usepackage{dcolumn}
\usepackage{bm}
\usepackage{verbatim}       

\usepackage[margin=4.3cm]{geometry}

\usepackage[dvips]{graphicx}
\usepackage{cite}
\usepackage{amsthm}

\usepackage{bm}
\usepackage{amstext}
\usepackage{amsmath}
\usepackage{amsfonts}
\usepackage{cite}

\newcommand{\dd}{{\rm d}}

\newcommand{\bd}{\begin{definition}}                
\newcommand{\ed}{\end{definition}}                  
\newcommand{\bc}{\begin{corollary}}                 
\newcommand{\ec}{\end{corollary}}                   
\newcommand{\bl}{\begin{lemma}}                     
\newcommand{\el}{\end{lemma}}                       
\newcommand{\bp}{\begin{proposition}}            
\newcommand{\ep}{\end{proposition}}                
\newcommand{\bere}{\begin{remark}}                  
\newcommand{\ere}{\end{remark}}                     

\newcommand{\bt}{\begin{theorem}}
\newcommand{\et}{\end{theorem}}

\newcommand{\be}{\begin{equation}}
\newcommand{\ee}{\end{equation}}

\newcommand{\bit}{\begin{itemize}}
\newcommand{\eit}{\end{itemize}}
\newtheorem{theorem}{Theorem}[section]
\newtheorem{corollary}[theorem]{Corollary}
\newtheorem{lemma}[theorem]{Lemma}
\newtheorem{proposition}[theorem]{Proposition}
\theoremstyle{definition}
\newtheorem{definition}[theorem]{Definition}
\theoremstyle{remark}
\newtheorem{remark}[theorem]{Remark}

\begin{document}


\title{Special relativity as the limit of an Aristotelian universal friction theory under Reye's assumption}

\author{E. Minguzzi\thanks{
Dipartimento di Matematica Applicata ``G. Sansone'', Universit\`a
degli Studi di Firenze, Via S. Marta 3,  I-50139 Firenze, Italy.
E-mail: ettore.minguzzi@unifi.it} }

\date{}
\maketitle

\begin{abstract}
\noindent
This work explores a classical mechanical theory under  two further assumptions: (a) there is a universal dry friction force (Aristotelian mechanics), and (b) the variation of the mass of a body due to wear is proportional to the work done by the friction force on the body (Reye's hypothesis). It is shown that mass depends on velocity as in Special Relativity, and that the velocity is constant for a particular characteristic value. In the limit of vanishing friction the theory   satisfies a relativity principle as bodies do not decelerate and, therefore, the absolute frame becomes unobservable. However,  the limit theory is not Newtonian mechanics, with its Galilei group symmetry, but rather Special Relativity. This result suggests to regard Special Relativity as the limit of a theory presenting universal friction and exchange of mass-energy with a reservoir (vacuum). Thus, quite surprisingly, Special Relativity follows from the absolute space (ether) concept and could have been
discovered following studies of Aristotelian mechanics and friction. We end the work confronting the full theory with observations. It predicts the Hubble law through tired light, and hence it is incompatible with supernova light curves unless both mechanisms of tired light (locally) and universe expansion (non-locally) are at work. It also nicely accounts for some challenging numerical coincidences involving phenomena under low acceleration.
%
%


\end{abstract}

\section{Introduction}

The Galilean principle of relativity establishes that the
mechanical laws are  the same for a family of observers in uniform
relative motion: the so called inertial observers. This important
principle, generalized by Einstein to all the physical laws,
provided, together with the principle of constancy of the speed of
light, the logical foundation of the special theory of relativity.

Already before this remarkable accomplishment, the relativity
principle was regarded as fundamental as it meant a radical
departure from the old Aristotelian physics according to which a
body, not acted upon by a force, would stay at rest in a privileged
absolute frame. To recognize the principle of relativity means to
recognize that there is no absolute space and that uniform motion
has to be understood as  {\em relative} to another observer.

Most modern physicists and philosophers would regard Aristotelian
physics as quite naive. It seems that Aristotle misinterpreted the
tendency of objects of coming at rest (with respect to the earth) as
the consequence of a general principle according to which a body
tends to its {\em natural state}, instead as the result of friction.
According to this school of thought the main merit of Galilean
physics is that of identifying the natural state of a body in
absence of friction as the fundamental one. This state is that of
uniform motion.

However,   this is not a fair historical reconstruction. As Dugas
\cite{dugas55} points out, Aristotle was very well aware that
probably, in vacuum, a body would have moved in uniform motion
indefinitely. However, from that he inferred that absolute vacuum is
impossible. In other words, it is not because Aristotle did not
recognize the role of friction that he did not come to the principle
or relativity, but rather, because he regarded the principle that
every body should come to absolute rest as more fundamental that the
relativity principle! Indeed, some kind of deceleration is necessary
in order to give physical observability to the concept of absolute
space which Aristotle was not prepared to abandon.


Despite this historical clarification, we can safely regard the
relativity principle as a cornerstone of modern physics, as we owe
to it the full development of Newtonian mechanics and the discovery
of Special Relativity. Nevertheless, the idea of cosmological flow
 and the discovery of the CMB radiation suggest that,  perhaps,
there is indeed a privileged reference frame and that, after all,
some elements of Aristotle's absolute space could still reenter into
play.

The aim of this work is twofold. First we will challenge most naive
criticisms to the  absolute space idea, showing that, in fact, some
aspects of relativity theory, like the relativistic mass formula,
are a natural consequence of Aristotelian mechanics. This rather
puzzling  result will be of interest  for the philosopher and the
physicist alike. Indeed, not even Galilean physics, which embodies
the relativity principle can claim a similar prediction (as it
misses the invariance of the speed of light). This result is based
on an assumption concerning the way in which friction alters the
mass (energy) content of a body. This is Reye's assumption, an old
relation from nineteen century applied mechanics which is promoted
here to a general principle. As we shall see, the role of the speed of
light will be played by a characteristic velocity which turns out to
be insensitive to friction. A body with this speed would
indefinitely preserve it. On the contrary, bodies having subluminal
speeds would decrease their velocity reaching a status of absolute
rest.
The just mentioned results will be independent of several details
concerning the universal friction force acting on bodies.

As a second objective we will explore the consequences of the simple
model of Coulomb (dry) friction. We will show that it predicts the
Pioneer anomaly and the Hubble law, where the latter is explained
through a kind of universal tired light mechanism. More importantly,
the model naturally explains the observed coincidence between the
Hubble constant and the Pioneer anomalous acceleration.

Then, we  show that the friction force can be given a perpendicular
component which is able to reproduce several aspects of MOND theory,
including the mechanism explaining the flatness of the rotational
curves of galaxies, the Tully-Fisher law, and the value of the
coefficient in the Tully-Fisher law. In other words, this model
seems to be able to explain many  odd phenomena that have been
observed for accelerations smaller or of the order of a certain
critical value $a_p$ although  it is not able to explain other observations like  the supernova light curves. Still, it could suggest some important ideas  towards the resolution of some puzzles of modern physics.

\section{Aristotelian mechanics}

Before we embark in these developments it will be convenient to
clarify what we mean by Aristotelian mechanics. There is a certain
consensus among historians that Aristotle had developed a {\em
Physics} rather than a {\em Mechanics}. His physics included
principles that would be hardly considered as mechanical by a modern
physicist but that allowed him to infer some mechanical
consequences. Certainly,  he had not in mind a simple axiomatic
structure as in the posterior Newtonian mechanics. Despite that,
there are repeated statements in his treatises which can be
converted into quantitative mechanical laws \cite{drabkin38}. Thus,
we can still imagine what an Aristotelian mechanics would look like
were it presented in mathematical language. Let us formulate
it through the following laws which will simplify the comparison with
Newtonian mechanics:
\begin{itemize}
\item[I] law: a body not acted apon by a force stays at absolute
rest,
\item[II] law: $m {\bf v}= \frac{1}{h} {\bf F}$,
\item[III] law: if a body $A$ acts with force ${\bf F}$ on a body $B$
then $B$ acts with force $-{\bf F}$ on $A$,
where these forces have the same line of action.
\end{itemize}
Here $h>0$ is a proportionality constant independent of the body and
with the dimension of a frequency. The proportionality constant $h$
could be omitted provided we redefined the unit of force. We shall
include an additional hypothesis which is also tacitly assumed in
Newtonian mechanics

\begin{itemize}
\item[Zeroth] law: Every body has a mass $m$ which is additive
(extensive). The total mass of \\  ${}$ \quad \   \ an isolated
system is preserved in time.
\end{itemize}

As it happens for Newtonian mechanics the first law, rather than
being a trivial consequence of the second, serves to remind us what
is the kinematics of the theory, namely what is the spacetime
structure. Indeed, the first law clarifies that there is a special
frame, and hence that spacetime has to be regarded as a product
between a real line of time $\mathbb{R}$, and an Euclidean space
$\mathbb{E}^3$. It is with respect to the Cartesian coordinates of
this frame that the second law is expressed.

The main information of the second law
stays in what is not explicitly stated, namely in the fact that
forces belong to a vector space and hence, that they can be added as
vectors. It also clarifies that, contrary to Newtonian mechanics,
Aristotelian mechanics is a first order theory. This should not come
as a surprise. For Aristotle the velocity of a body is proportional
to the force applied on it, and inversely proportional to its mass
(Physics VII 249b-250a). As it has been observed by some authors,
the analog of Newton's second law for Aristotle is then Stokes' law.
The acceleration phase is purely transitory and due to the fact that
the applied force could be variable. Admittedly, this theory cannot explain the  approximate uniform motion of a projectile
without invoking some weird mechanism to justify its slow
deceleration. Indeed, in the middle ages the example of the
projectile was often used by J. Buridan and other philosophers to
show that Aristotelian mechanics was untenable.

We included a third law, coincident in form with that of Newtonian
mechanics. It has some desirable consequences. For instance, through
it it is possible to prove that the center of mass of an isolated
system does not change in time. The Newtonian version states that it
moves with uniform velocity but here it must correctly keep the same
position in the absolute frame since by isolation no exterior force
can set it in motion. Thanks to the second statement of the  third
law we have  that the angular momentum of any isolated system
vanishes. Thus, without some exterior force a rigid
macroscopic body cannot neither translate nor rotate.

In what follows  with {\em Aristotelian mechanics} we shall
generically refer to a broader set of theories which we now
introduce. We take for well established that a correct mechanics
should be a second order theory, and that the first order version
given above should be recovered ignoring the transitory acceleration
phases. We replace the previous laws with the following
\begin{itemize}
\item[I'] law: A body,  not
acted on by a force, decelerates towards its natural state of \\
${}$ \quad \   \ absolute rest,
\item[II'] law: $m {\bf a}=  {\bf F}_f+{\bf F}$,
\item[III] law: if a body $A$ acts with force ${\bf F}$ on a body $B$
then $B$ acts with force $-{\bf F}$ on $A$,
where these forces have the same line of action,
\end{itemize}
plus the Zeroth law.  Here ${\bf F}_f$ is a universal friction force
which depends on the velocity of the body,  on its mass and, in a
special version that we shall explore, even on the external force
${\bf F}$. We shall assume that at least the force component
parallel to the velocity, and hence responsible for the negative
work, be proportional to mass
\begin{equation} \label{soo}
{\bf F}_f=-a^{\parallel}_f m \hat{{\bf v}}+{\bf F}_f^{\perp}
\end{equation}
where $a^{\parallel}_f$ might depend on velocity. Furthermore, we
shall
 assume that in absence of other forces ${\bf F}_f$ has
no perpendicular component. The force to which the first and third
law refer to is ${\bf F}$, since it is understood that the friction
force acts on every body. The reader might assume  ${\bf F}_f^{\perp}= 0$ on first reading since our interest in the general case will be clarified only in Sect.\ \ref{map}.

The above classical form of Aristotelian mechanics is then recovered
for \[{\bf F}_f=-h m {\bf v}.\] This linear dependence on momenta is
particularly useful as it shows that the second law can be assumed
for imaginary point particles, the same law for macroscopic bodies
being just a consequence of linearity (through the usage of the
center of mass). In particular, other options for ${\bf F}_f$ should
come with a specification of what {\em body} really means in the
previous laws, as the non-linearity of the theory would imply a
failure of the extensive property.

We end the section remarking the coherence of
Aristotelian mechanics. Indeed, it is never sufficiently stressed
that there is an nice consistency between the first and second laws
and the spacetime product structure $\mathbb{R}\times \mathbb{E}^3$.
Only the assumption of this absolute spacetime structure allows us
to make sense of the first and second law, and conversely, without a
friction force the theory would reduce to Galilean mechanics and
the absolute space would become unobservable and would have
no clear epistemological status.

\section{Reye's assumption and relativistic mass}

As the above formulation clarifies, Aristotelian mechanics is
nothing but Newtonian mechanics plus a friction force. Curiously, we
 do not have  to change the spacetime structure since Newton formulated
his theory on absolute space, though the lack of a
universal friction force prevented its identification.
  In modern physics there is  a way of
expressing Newtonian physics without resorting to the concept of
absolute space, namely using the concept of fibration over time. The reader is referred to \cite{penrose68,ehlers73,trautman70,heller06} \cite[Chap.\ 17]{penrose05}.

In the framework of classical mechanics let us consider the motion
of a body subject to a friction force ${\bf F}_f$. Later we shall
consider the introduction of an additional force ${\bf F}$. We wish
to take into account the effect of friction on mass. In order to fix
the ideas the reader might think of a block moving on a rough
surface. We assume that the block loses mass because of wear, and
that in a given time interval the lost mass (equal to the debris
mass) is proportional to the work done by friction forces.

This is
the Reye's hypothesis for dry friction\footnote{Historical note. In German and Italian University courses
in applied mechanics Reye's assumption has been  taught at least since the first half of the 20th century \cite{bach01,funaioli73,villaggio01}. Though applications of this theory appeared in English  \cite{opatowski42}, Reye's  ideas were long ignored in the English and American literature. Similar conclusions have been  reached only much later by other authors \cite{holm46,archard53}. Reye's assumption allows one to calculate the distribution of pressure in the contact of two surfaces and hence to extract the friction force, e.g. a rotating horizontal disc above a horizontal plane.}
 \cite{reye60}. This assumption is simple and elegant because it
basically says that the work done by friction forces, rather than being completely dispersed into heat, goes in a given proportion into the  breaking
of the molecular bonds that keep the block molecules together.
Although called ``assumption'' or ``hypothesis'' this is really an experimental fact in its own domain of applicability. As we shall promote it to a universal law, we shall apply it   at regimes of velocity, mass, and acceleration
which go far beyond the framework of applied mechanics that
originally motivated it. Mathematically, it reads
\[
\dot m=\frac{1}{c^2} \,{\bf F}_f\cdot {\bf v} , \qquad (Reye)
\]
where $\frac{1}{c^2}$ is Reye's proportionality constant where $c\in (0,+\infty]$ has  the dimension of a
velocity.
Let $F^{\parallel}_f$ and $F^{\perp}_f$ be, respectively, the module
of the component of the friction force parallel to ${\bf v}$, and
perpendicular to ${\bf v}$. Using Eq.\ (\ref{soo}) we obtain
\begin{equation}
\dot{m}=-\frac{1}{c^2}\, a^{\parallel}_f p.
\end{equation}%
%
The idea behind this Reye's type stipulation is that the friction force arising
in our model will be due to the interaction with a pervasive {\em
vacuum}, alternatively called {\em reservoir}. If this law could be
proved to be true then, given more information on the nature of the
vacuum state, it could possibly be justified with some kind of
microscopic mechanism. However, at this stage we do not try to make
 assumptions on the nature of this medium and take the above law
as given.

It remains to write down the first cardinal equation for the motion
of the body under friction forces. To fix the ideas the reader might still think at the  example of the block moving
on a rough horizontal surface. Since we are in
presence of a variable mass system we have to use the formula
\begin{equation} \label{nhi}
m \dot{\bf v}= {\bf F}_f-\dot{m} ({\bf v}-{\bf v}_d)
\end{equation}
where ${\bf v}_d$ is the velocity of the debris. This formula much
used in rocket theory goes back to Painlev\'e and  Seeliger (1890)
and can be deduced from the conservation of momentum (see
\cite{irschik04} for a nice account on the history of this formula).
Since the debris have been detached because of their motion with
respect to the vacuum (the horizontal surface in the block example) it is natural to assume that after
detachment the debris do not move anymore with respect to it (i.e.\ they become part of the vacuum), thus
${\bf v}_d={\bf 0}$,
\begin{equation} \label{vhq}
m \dot{\bf v}= {\bf F}_f-\dot{m} {\bf v}
\end{equation}
Denoting with ${\bf p}:=m {\bf v}$ the linear momentum we obtain the
system of equations
\begin{align}
\dot{m}&=-\frac{1}{c^2}\, a^{\parallel}_f p,  \label{jhg}\\
\dot{{\bf p}}&= {\bf F}_f=-a^{\parallel}_f m \hat{{\bf v}}. \label{jhh}
\end{align}
These are the equations that govern the motion of a free body in our
theory. We regard equation (\ref{jhg}) as a consequence of Reye's
assumption, and hence as the result of the balance between work and
mass transfer. Equation (\ref{jhh}) must instead be regarded as the
consequence of an instantaneous conservation of momentum in the
interaction with the vacuum state.
A special solution is provided by $v=0$,
according to which mass is constant. Bodies which satisfy this
condition are said to be at {\em absolute rest} or simply at rest.

Multiplying the first equation by $m$, multiplying scalarly the
second equation by $\frac{1}{c^2}\, {\bf p}$, and taking the
difference we obtain that there is a constant $m_0$ such that
\begin{equation} \label{kos}
m^2 -\frac{1}{c^2} {\bf p}^2 =m_0^2 ,
\end{equation}
or equivalently
\begin{align}
m&=\frac{m_0}{\sqrt{1-\frac{v^2}{c^2}}}, &v<c   \qquad \ \quad \qquad if \ m_0>0, \label{ngx}\\
mc^2&=p \,c,  &v=c, \qquad \ \quad \qquad if \
m_0=0, \label{ngy}\\
m&=\frac{\vert m_0\vert }{\sqrt{\frac{v^2}{c^2}-1}}, &v>c \qquad if
\ m_0 \textrm{ is imaginary}, \label{ngz}
\end{align}
Thus both mass and velocity are related as in Special Relativity (the
latter being the tachyonic case). The time dependence is
nevertheless different, as in the first bradyonic case both velocity
and mass decrease while preserving the relativistic relationship.
From Eq.\ (\ref{jhg}) and (\ref{jhh}) taking into account that the
mass $m$ is always positive, we obtain
\begin{equation} \label{ndf}
\dot{\bf v}= ({\bf F}_f-\dot{m} {\bf v})/m=-{a^{\parallel}_f}
(1-\frac{v^2}{c^2})\hat{{\bf v}}.
\end{equation}
This equation  shows that any body with velocity $c$ will keep
moving in straight line preserving it. In other words, a body moving
at this characteristic speed is insensitive to the deceleration
which one would intuitively expect by friction. Of course, in the example of the block one should not expect the existence of such a special velocity as $c$ would be very high at a range of velocities for which Reye's assumption does not hold.



Even bodies admitting the special velocity $c$ will experience a
non-trivial dependence of mass on time.  From Eq.\ (\ref{jhg}) we
obtain
\begin{equation} \label{bsc}
m(t)=m(0)\, e^{-\frac{1}{c}  a_f^\parallel(c) t}, \qquad \textrm{ for }
v=c.
\end{equation}
Thus it decreases since  $a^\parallel_f>0$.

In the bradyonic case Eq.\ (\ref{ndf}) implies that the velocity will
further decrease preserving the bradyonic condition $v<c$. Given the
dependence of mass on velocity  (\ref{ngx}), as the velocity goes
to zero the mass $m$ approaches $m_0$. We can call $m_0$ the {\em
absolute rest mass} where we added the adjective {\em absolute} to
avoid confusion with the special relativistic interpretation. Here
$m_0$ is the mass as measured in the absolute frame when the
particle has (approximately) come to rest. So far we provided no
connection with the mass as measured in a frame comoving with the
body. We shall call $m_0$  {\em absolute rest mass} even for
$m_0=0$ though in this case the terminology is improper as the body
will proceed at constant speed without ever reaching a state of
absolute rest.

Similar considerations prove that a body in a tachyonic status
increases its velocity thus preserving that condition. The mass
decreases and goes to zero according to Eq.\ (\ref{jhg}).

We now consider two possible models for the parallel component of
the friction force. From section \ref{lim} we shall only consider
the Coulomb case as it is favored by some cosmological and astronomical data. It is also more natural given the origin of the theory, since Reye's assumption was conceived for dry friction for which the Coulomb's type force provides the best approximation.
 We
shall also restrict most of the analysis to the case $v\le c$.

\subsection{Coulomb friction} \label{msx}
Let us calculate the dependence of velocity on time for Coulomb
friction
\begin{equation}
{\bf F}_f=-m a_p \hat {\bf v},
\end{equation}
where $a_p$ is a characteristic acceleration (for the example of the
block over a rough horizontal surface $a_p=\mu g$ where $g$ is the
gravitational acceleration and $\mu$ is the dynamic friction
coefficient). The subscript $p$ stands for {\em parallel}. We have
$a^\parallel_f=a_p$ and from Eq. (\ref{ndf}) we get
\begin{equation}
\dot{\bf v}=- a_p (1-\frac{v^2}{c^2}) \hat {\bf v}.
\end{equation}
Thus, either $v=0$ or
\[
\dot{v}=-a_p(1-\frac{v^2}{c^2}).
\]
This equation can be integrated and gives
\begin{align*}
v(t)&=c \tanh (\varphi-\frac{a_p}{c} \,t), & \textrm{for} \ 0\le v<c\\
v(t)&= c \tanh^{-1} (\varphi-\frac{a_p}{c} \,t), & \textrm{for} \ v>c.
\end{align*}
where $\varphi$ is an integration constant (interpreted as the
rapidity in the former case). We observe that in the bradyonic case,
and for Coulomb friction, the time needed to reach a complete stop
is finite, namely $\Delta t=c \varphi/a_p$.

The dependence of mass on time is
\begin{align*}
m(t)&=m_0 \cosh (\varphi-\frac{a_p}{c}\, t), & \textrm{for bradyonic  mode}, \\
m(t)&= \vert m_0\vert \sinh (\varphi-\frac{a_p}{c} \,t), & \textrm{for
tachyonic mode}.
\end{align*}
In the bradyonic case it becomes equal to the absolute rest mass
when the particle reaches absolute rest. In the tachyonic case it
goes to zero quite rapidly as it happens for the lightlike case.

If a body moves at speed $c$ then its mass decreases as: $m(t)=m(0)\,
e^{-\frac{a_p}{c}\, t}.$

\subsection{Stokes friction}
Let us calculate the dependence of velocity on time for
friction forces proportional to velocity
\begin{equation}
{\bf F}_f=- h m  {\bf v},
\end{equation}
where $h$ is a constant with the dimension of a frequency. We have
$a^\parallel_f=h v$ and from Eq. (\ref{ndf}) we get that ${\bf v}$ has constant direction and
\[
\dot{v}=-h(1-\frac{v^2}{c^2})v.
\]
This equation can be integrated and gives
\[
\frac{\vert 1-\frac{v}{c}\vert }{1+\frac{v}{c}}
(\frac{v}{c})^2=Ke^{-2ht},
\]
where $K\ge 0$ is an integration constant. This equation could be
inverted to find $v(t)$ but the analytic expression is not
particularly illuminating. What is important is that for small
velocities $v\propto e^{-ht}$ thus though the body will take
infinite time to come to rest, it will cover a finite path and
hence, at any practical effect, it can be considered to come to rest
in a finite time, namely when the amount of space to be covered
becomes negligible with respect to the size of the body.

The dependence of mass on time can be deduced from that of the
velocity. Indeed, using Eq.\ (\ref{jhh}) we obtain that they are
simply related through the equation $mv=K' e^{-h t}$. It is convenient to define $a_p:=h c$
so a body which moves at speed $c$ has its mass decrease as $m(t)=m(0)\,
e^{-\frac{a_p}{c}\, t}$ as for Coulomb friction.

\section{Limits and completion of the theory} \label{lim}
From this section we consider only the Coulomb friction case, although a similar analysis could be performed for $a_f^\parallel$ dependent on velocity. For instance, in order to deal with Stokes friction it is sufficient to replace `Coulomb' with `Stokes' in the next instances.

In the previous study there appeared two constants. The velocity $c$
and the acceleration $a_p$. In the limit $c \to +\infty$ the
previous equations show that the mass does not change, and hence the
theory describes simply a body moving under Coulomb friction in the
simplest mechanical case which does not introduce Reye's hypothesis.
Thus for low speed the theory reduces to Aristotelian mechanics with Coulomb friction.

We wish to show that in the limit $a_p \to 0$ the theory gives
Special Relativity, or equivalently that for large accelerations the theory reduces to Special Relativity. It must be remarked that so far we have made no
assumptions which relate observations made in the absolute frame with observations performed on a
frame moving uniformly with respect to it.
Therefore, the theory is incomplete and must be complemented with
further assumptions to be contrasted with observations.

In this respect  it is useful to realize that the mere symmetry of
the dynamical equations with respect to some group of coordinate
transformations, as for Newton laws with respect to Galilei
transformations, does not imply a relativity principle. In fact, one
of the physical contents of the relativity principle is precisely
that of clarifying that the new coordinates are not mere artifacts,
but are instead connected with measured lengths and time intervals
in a different frame. Thus, the validity of a relativity principle,
though permitted by the formalism, is ultimately a subject of
observation.

Coming to our model, in the limit $a_p \to 0$ the friction vanishes,
the natural motion for free particles becomes the uniform motion,
and it becomes impossible, in our Aristotelian theory complemented
with Reye's assumption, to observe the underlying absolute space. As
a consequence, the theory admits the mathematical possibility of
embodying, in this limit, a relativity principle. We therefore
stipulate that the theory should predict the validity of a
relativity principle in the limit $a_p \to 0$. In fact, we must
stipulate something more if we want to connect the observations made
in different frames for $a_p\ne 0$. It is natural to embody in our
theory the following weak form of the relativity principle
\begin{quote}
Every law of physics which does not involve in any non-negligible
way the universal friction force ${\bf F}_f$, must be invariant for
the set of (inertial) observers which move uniformly with respect to
the absolute reference frame.
\end{quote}
Since the effects of the friction force become negligible for
$a_p\to 0$, this weak relativity principle implies the usual
relativity principle in that limit.

There are very few possibilities concerning the possible symmetry
groups that could express the relativity principle for $a_p\to 0$.
Under mild conditions on causality, parity, homogeneity and isotropy
they reduce to the Galilei and Lorentz groups
\cite{levyleblond76,levyleblond79}.

The theory developed so far naturally suggests which  group should
be used. Indeed, let us observe that in the limit $ a_p \to 0$
equation (\ref{ngx})
\[
m(v)=\frac{m_0}{\sqrt{1-v^2/c^2}},
\]
expressing the dependence of mass on velocity, and Eq.  (\ref{kos})
 do not change, while equations (\ref{jhg})-(\ref{jhh}) reduce to the
correct ones for free motion in Special Relativity.

It is known that an extensive quantity $m$ in a theory which admits
the Lorentz relativity group must depend on velocity as follows
\[
m(v)=\frac{m_0'}{\sqrt{1-v^2/{c'}^2}},
\]
where $m_0'$ is the rest mass of the body (i.e. measured in its
comoving frame) and $c'$ is the constant speed of particles with
$m_0'=0$, while it must be independent of velocity for a Galilei
relativity group, i.e. $m(v)=cnst$ \cite{penrose65d,ehlers65}.

In our theory with $a_p\to 0$ (think of $a_p$ as arbitrarily small)
let us consider  the slow deceleration of the body which reaches
absolute rest in the remote future. Our expression for $m(v)$ must
coincide with that deduced from the relativity principle (the
coincidence for two different values of $v$ will suffice). In
particular, we have to discard the Galilean possibility and
conclude that the relativity group for $a_p\to 0$ has to be the
Lorentz group and that $m_0=m_0'$, $c=c'$. This argument proves that
$m_0$ is not only  the {\em absolute rest mass} in the remote
infinite future  but also the {\em rest mass} as measured in the
comoving frame (at least for $a_p\to 0$).

This conclusion does not change for $a_p\ne 0$, unless we complicate
our theory with the introduction of additional characteristic
dimensional quantities. Indeed, the dependence of $m_0'$ can be
expressed $m_0'=m_0 f(a_p,m_0,c)$ where $f\to 1$ for $a_p\to 0$. For
dimensional reasons the only dimensionless function is $f=1$, and
hence we conclude that $m_0=m_0'$ even far from the limit $a_p\to
0$.

Although we know that the theory does not respect the usual
relativity principle - for it suffice to wait enough to observe in
which privileged  frame all the bodies go at rest - we know, by our
weak relativity principle, that it must satisfy a relativity
principle for all phenomena that do not involve the universal
friction. For this restricted set of phenomena the kinematical
symmetry group is again the Lorentz group for, essentially, the same
argument given above. One can obtain this result from another route,
namely taking into account that Landau and Sampanthar
\cite{landau62} proved that if mass is extensive, conserved, depends
on velocity and on rest mass as we just established, and the
relativity principle holds, then the relativity group of
transformation is the Lorentz group.

%
%

\begin{figure}[ht!]
\centering
 \includegraphics[width=7cm]{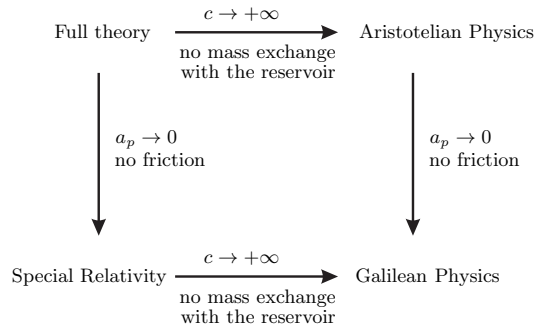}
 \caption{The proposed full theory, a Aristotelian theory under Coulomb friction and Reye's hypothesis, unifies two regimes, that of Aristotelian physics, obtained for small velocities, and that of Special Relativity, obtained for large accelerations.}
\label{pr}
\end{figure}

We are therefore lead to the conclusion that our theory coincides
with Special Relativity in all aspects that do not involve the
universal friction, including time dilation and mass-energy
equivalence. Mathematically, it is coincident with Special Relativity in which we added an acceleration field on the
relativistic velocity space $\mathbb{H}^3$ - a field which takes a
particular symmetric form in  the privileged absolute frame. Of
course, the interpretation is quite different.

We wish to stress that this natural conclusion has been obtained
without using the frame invariance of the speed of particles with
$m_0=0$. Indeed, we started from a universal friction theory and
concluded that the limiting theory for $a_p\to 0$ should obey a
relativity principle, and should in fact be Special Relativity with
its Lorentz group symmetry, rather than classical mechanics with its
Galilei group symmetry. Our  argument clarifies
the tight link between relativity theory and friction. In particular, it shows that the  absolute space (ether) concept might directly lead to Special Relativity, a conclusion which is at odds with common knowledge.

Following the above thread of arguments there was indeed the
possibility of obtaining a Galilean relativity group and hence
Newtonian mechanics. We had to switch off Reye's assumption placing
$\frac{1}{c}=0$. However, since this condition is equivalent to
$c\to +\infty$ in our model, we have that this possibility is merely
a special case of our theory according to the  usual limit by which
relativistic physics reduces to classical physics for low speeds.

Finally, we wish to comment on the epistemological status of the
relativistic mass concept. There have been repeated objections
towards the introduction of this concept in Special Relativity on
the ground that the concept of mass should not be relative to a
frame. According to this school the concept of relativistic mass
should be avoided in favor of the concept of energy
\cite{adler87,okun09}. While I am partially sympathetic with these
ideas, we have to admit that the concept of relativistic mass is the
one that preserves the addictive and extensive properties of the
mass concept in classical mechanics, and which can therefore serve
the intuition in the transition from classical to relativistic
physics \cite{sandin91,jammer00}. Whether this intuition is really
misleading is still a subject of debate. However, I remark that in
the present theory the mass $m(v)$ has a completely different and
more objective epistemological status. For, if we restrict the usage
of $m(v)$ with respect to the absolute frame, then the concept
becomes absolute rather than relative. Clearly, the whole
interpretation of our theory is based on this mass, and in this
section we learned that it can be reinterpreted as energy up to the
usual $c^2$ factor: $E=m c^2$.

Thus in our theory, the equations of conservation of mass and linear
momentum  for a collision e.g. ($i$ stands for {\em initial} and $f$
for {\em final})
\begin{align}
m_{1i}+ m_{2i}&= m_{1f}+ m_{2f} ,\\
 {\bf p}_{1i}+ {\bf p}_{2i}&= {\bf p}_{1f}+ {\bf p}_{2f}.
\end{align}
written down for the absolute frame, are such that the former can be
equally well interpreted as a mass conservation equation (Zeroth
law) or as a relativistic energy conservation equation.

A final word must be spent to properly account for particles with vanishing rest mass $m_0=0$.
Equation (\ref{kos}) or (\ref{ngy}) show
that there is a mass (energy) indeterminacy connected with this type
of particles. In order to fix this indeterminacy it is necessary to
assume that some observable quantity relates with it. As it is done
in Special Relativity, we identify particles moving at speed $c$ with
photons and assume Planck's formula
\begin{equation}
E=h_P \nu, \qquad \textrm{ if } v=c,
\end{equation}
where $\nu$ is the frequency of the photon and $h_P$ is the Planck constant.

\section{Additional forces} \label{map}

Given the established equivalence between mass and energy it is
convenient to rewrite Eqs. (\ref{jhg})-(\ref{jhh}) through the
variable $E=mc^2$ (observe that ${\bf p}=m{\bf v}$ implies ${\bf
p}c^2=E {\bf v}$).
\begin{align}
\dot{E}&=- a_p p,  \label{jhg2}\\
\dot{{\bf p}}&= -\frac{a_p E}{c^2} \, \hat {\bf p} +{\bf F}^\perp_f.
\label{jhh2}
\end{align}
If we include an additional force, taking into account the effect of
work on energy
\begin{align}
\dot{E}&=- a_p p+{\bf F}\cdot {\bf v},  \label{jhg3}\\
\dot{{\bf p}}&= -\frac{a_p E}{c^2}\, \hat {\bf p}+{\bf
F}^\perp_f+{\bf F}. \label{jhh3}
\end{align}
For a vanishing universal friction (i.e.\ $a_p=0$, $F^\perp_f=0$)
these reduce to the usual formulas in Special Relativity, and imply
the conservation of the scalar: $m_0^2c^4=E^2-p^2c^2$.

In must be recalled here that the force in Special Relativity is
conveniently defined as  $\dot{{\bf p}}$, because in this way
the conservation of momentum implies Newton's third law, and
moreover, from $E=c\sqrt{{ \bf p}^2+m_0^2c^2}$ it follows
$\dot{E}={\bf F}_{tot}\cdot {\bf v}$, which is the usual kinetic
energy theorem. The so called covariant world force $F^\mu:=\dd
p^\mu/\dd \tau$, on the contrary, is of little utility especially
when considering collisions.

Equation (\ref{jhg3}) can still be interpreted as a Reye's type of
relation. Indeed, it can be rewritten
\[
\dot{m}=\frac{1}{c^2} \dot{L}
\]
where $\dot{L}=-a_p m v+{\bf F}\cdot {\bf v}$ is the work done in
the unit of time by the friction force and by the other forces.

Thus,  the force ${\bf F}$ so embodied in the theory respects the
relativity principle (as the preservation of $m_0$ clarifies). This
fact can be interpreted by saying that forces implemented in this
way are unrelated to the friction that brakes the symmetry, but not
to the vacuum and to Reye's relation. In fact Reye's relation holds
for them too, pointing to the idea that all interactions are
mediated by the vacuum and should not be thought as pertaining to
the interacting bodies alone. This observation might be a
suggestions on how to implement this theory in a field theoretical
way.

Decomposed ${\bf F}={ F}^\parallel \hat{\bf p}+{\bf F}^\perp$, and
using $\dot{\bf v}=\dot v \hat{\bf v} +v\, \dot{\hat{{\bf v}}}=\dot
v \hat{\bf v}+{\bf a}^\perp$, we can write Eq. (\ref{jhh3}) as the
system
\begin{align}
\dot{p}&=-\frac{a_p E}{c^2}+F^\parallel, \label{puo}\\
m {\bf a}^\perp&={\bf F}^\perp_f+{\bf F}^\perp. \label{pun}
\end{align}

Equation (\ref{puo}) is quite interesting because it proves that no
bradyonic particle can exceed the characteristic speed $c$. Indeed,
since $m_0$ stays constant, Eq. (\ref{puo}) holds true as long as
$v<c$. This equation shows that $m(v)$ goes to infinity for $v\to c$
hence the first friction term in the right-hand side of Eq.
(\ref{puo}) becomes greater than $F^\parallel \hat {\bf p}$ causing
the velocity to stop its growth at a value $v_{max}$ for which the
right-hand side of Eq. (\ref{puo}) vanishes, namely
\begin{equation} \label{scp}
v_{max}=c \sqrt{1-(\frac{m_0a_p}{F^\parallel})^2}, \qquad \textrm{
for } F^\parallel>m_0a_p.
\end{equation}
At this velocity the mass becomes constant, namely $m=F^\parallel/a_p$.
Equation (\ref{scp}) is reminiscent of the original {\em second law of
Aristotelian dynamics}, namely the claim  that the velocity of a
body is proportional to the force acting on it. Indeed, in our
theory there is a monotonous increase of velocity with force, but
there is no linearity because we adopted  a Coulombian friction
force rather than a Stokes' type force. Since $m_0a_p/F^\parallel$
is very small, in any practical circumstance in which
$F^\parallel\ne 0$, $v_{max}$ is very large and it becomes very
difficult to observe the above relation between force and velocity.

The just given argument for the preservation of the bradyonic status
of particles works unaltered for impulsive forces, and hence for
collisions. Curiously, in Special Relativity textbooks it is very
common to find incorrect explanations for the fact that particles
with positive rest mass  cannot reach the speed of light. These
arguments, based on the dependence of the relativistic mass on
velocity, are simply fallacious as they provide a dynamical
mechanism for something that cannot happen already for kinematical
reasons (the very existence of the light cone and the very fact that
 massive particles are represented as timelike worldlines). On the contrary in our
theory, the kinematics does not demand that the light cone be
invariant (though this turned out to be the case). Thus the just
given dynamical explanation for the preservation of the bradyonic
status is acceptable.

Equation (\ref{puo}) faces us with a new problem. What happens when
the object is initially at rest (${\bf v}={\bf 0}$) and a force
$F^\parallel \le a_p m$ is applied? In order to start moving, say at
time $t=0$, we must have $\dot{p}(\epsilon)>0$  at some later instant. But
the right-hand side reads $-a_p m+F^\parallel < 0$. Thus the particle
does not move and moreover we must admit that since the left-hand
side vanishes, also the right hand side vanishes. As $F^\parallel\ne
0$, writing the vectorial version of the equation, this is possible only if $-a_p m\hat{\bf v}$ can be different
from $0$ for $v=0$. In particular it takes any value in a sphere of
radius $a_p m$, as long as that value allows to solve the
differential equation with the solution ${\bf v}=0$. We are
therefore led to the conclusion that it is necessary to introduce a
static friction coefficient. The radius $a_p m$ correspond to the
case $\mu_s=\mu_d$ but more generally it can have value $\mu_s\ge \mu_d$.

\section{Comparison with observations}

We shall consider the so far developed theory  as a local
approximation. Let us suppose that the Universe is a
patchwork of possibly overlapping and interacting vacuum states.
As an analogy, consider a table $V_0$, a sheet of paper $V_1$, and a coin
moving over $V_1$. The coin is the body which feels the friction of
$V_1$, which in turn can be in motion  responding to the friction of
$V_0$. Of course there could be different sheets on the table and different coins over the same sheet.

We shall suppose that there could be a vacuum state $V_S$ at the
level of the solar system, which in turn behaves as a body with
respect to the vacuum state $V_G$ extending all over the Milky Way,
which in turn moves over a vacuum  state extending all over the
local group and so on. We might consider that due to friction, two
vacuum states will tend to be at relative rest, and when this happen
they will form a single vacuum. We shall not consider the details of
this interaction and the way by which one body ends interacting with
a vacuum instead of another.

\subsection{Pioneer anomaly}
A first obvious consequence of our model is that free bodies with small velocities compared to $c$ should
show a deceleration of magnitude $a_p$ in the absolute reference
frame (Sect.\ \ref{msx}). The effects of the universal friction should become
measurable when $F^\parallel/m\sim a_p$.

Let us consider the vacuum state at the level of the solar system
$V_S$. Any body should decelerate with an acceleration of magnitude
$a_p$. The Pioneer spacecrafts do present an unmodeled acceleration
approximately directed towards the Sun of magnitude
\cite{Anderson:2001sg}
\[
a_P=(8.74 \pm 1.33)\times  10^{-10} m/s^2.
\]
Unfortunately,  according to a recent study, the anomalous
acceleration seems to be non collinear with the spacecraft velocity
\cite{turyshev11}, and can be accounted for by thermal radiation
\cite{turyshev12}.

Also the friction model  predicts a deceleration of the planets of the
solar system, which decreases the radius of their orbits and hence
implies an increase in the modulus of their velocity \cite{parkyn58}
(in the end $\dot{v}=a_p$, that is, the sign that one would naively
expect gets inverted under Coulomb friction). However, this effect
is quite small and can be cancelled by the decrease of the Sun mass
due to solar wind \cite{krasinsky04}.

\subsection{Hubble law}
Let us now consider the vacuum states containing our local group of
galaxies and the local supercluster $V_{LS}$. Because of friction we
must admit that most galaxies are almost at rest. Nevertheless, by a
tired light effect the Hubble law still holds. Indeed, a photon sent
at time $t$ from a galaxy and received at time $t+\Delta t$ in the
Milky way undergoes a redshift (see Eq. (\ref{bsc}))
\[
1+z=\frac{E(t)}{E(t+\Delta t)}=e^{\frac{1}{c} \,a_p\, \Delta t}\simeq
1+ \frac{1}{c} \,a_p \,\Delta t.
\]
If this tired light explanation for the  Hubble law is correct then
we must find that the observed value for
\[
a_H:= cH=(6.9\pm 0.9) \times 10^{-10} m/s^2
\]
(if $H = (72\pm 8) (km/s)/Mpc$)  coincides with $a_p$, and if the
above explanation for the Pioneer anomaly is correct we must expect
$a_p=a_P$ and in the end
\[
a_H=a_P.
\]
The above figures seem to confirm this prediction. This equality has
been noticed by many authors \cite{Anderson:2001sg}, but this
appears to be the simplest physical theory which accounts for it. This explanation is the more striking as it is compatible with Special Relativity.

%
%

This is a kind of tired light explanation of the Hubble law, however, contrary to
usual tired light theories, it does not assume that the loss of energy is due to scattering with diffuse interstellar matter. The latter assumption would imply a modification in the direction of the photon, so that any galaxy would be seen as blurred and indefinite, contrary to observations. Our universal friction mechanism preserves the direction of the photon and so avoids this problem.

Unfortunately,  tired light explanations of the Hubble law cannot
account for the supernova light curves \cite{leibundgut96}.
As it will be clarified in the next subsections, perhaps the Hubble law could be due to two mechanisms, namely tired light for sufficiently close cosmological objects (with the idea that they belong to the same vacuum state) and cosmic expansion for far away objects (as they belong to different vacuums states, different states diverging from each other). In other words, in the empty region between two vacuum patches there would be no friction and so the dynamics would be completely relativistic (recall that the theory becomes relativistic for $a_p\to 0$, and by Eq.\ (\ref{bsc}) there is no tired light effect in that limit). There would still  be a redshift effect but due to the relative motion of the vacuum states. One would have to explain why these two redshift effects, namely that due to tired light and that due to the universe expansion have the same magnitude.


\subsection{Coldness of the Hubble flow}
If the Hubble law
is at least locally due  to tired light, then it should be expected to hold for close
celestial objects as well. The expanding universe theory predicts
the Hubble law at length scales which are well beyond the scale of
homogeneity for which a Friedmann-Robertson-Walker  approximation of the cosmological
metric would make sense. A well known puzzle in cosmology is the
``coldness of the Hubble flow'' namely the observation that the
Hubble law holds at the scale of the local group (1-10Mpc) with a
velocity scatter with respect to the Hubble flow which is very small
(40 km/s), although at that scale the matter distribution is very
clumped \cite{baryshev01}. A local tired light model for the Hubble law
accounts for this observation quite easily, for according to this
explanation the local Hubble law does not depend on an expansion dynamics.


\subsection{Perpendicular friction and MOND}

So far we have not specified the dependence of ${\bf F}^{\perp}_f$
on exterior data. We  consider the following model
\begin{equation}
{\bf F}^\perp_f=m a_0 \,\beta\Big(\frac{F^\perp}{ma_0}\Big) \, \hat {\bf F}^\perp, \qquad a_0:=q a_p
\end{equation}
where $q$ is a dimensionless number of the order of unity and $\beta: [0,+\infty)\to [0,+\infty)$ is a non-decreasing function such that $\beta(x)\sim \sqrt{x}$ for $x \ll 1$ and $\beta(x)/x\ll 1$ for for $x \gg 1$.
According to this stipulation the perpendicular component of the friction force
depends on the applied exterior force and becomes negligible for $a_p\to 0$. The assumption that $q$ is of the order of unity
essentially means that we are not introducing any other
dimensional parameter in our model.

Let us consider the motion of stars in a spiral galaxy under the
assumption that there is a vacuum $V_G$ extending all over it.
Equation (\ref{pun}) reads
\[
m  a^\perp=m a_0\, \beta\Big(\frac{F^\perp}{ma_0}\Big)+ F^\perp.
\]
where $m$ is the mass of the star, $a^\perp$ the magnitude of the
component of the  acceleration which is perpendicular to the
velocity, and $F^\perp$ is the magnitude of the exterior force
perpendicular to the velocity. The previous equation can be inverted
to give
\begin{equation} \label{lmo}
m a^\perp \mu(a^\perp/a_0)=F^\perp, \qquad a_0:=qa_p
\end{equation}
where $\mu$ is a function such that $\mu(x)\to 1$ for $x\gg 1$ and $\mu(x)\sim
x$ for $x \ll 1$. For instance, if $\beta=\sqrt{x}$ we have
\begin{equation}
\mu(x)=1-\frac{1}{2x}(\sqrt{1+4x}-1).
\end{equation}
We recognize in Eq.\ (\ref{lmo}) the  MOND relation \cite{milgrom83a,sanders02,famaey12}, at least for
what concerns the degrees of freedom perpendicular to the velocity.
Fortunately, they are the most important in the derivation of MOND
type phenomenology.

Indeed, let us recall how the Mond mechanism works applying the equation to the circular motion of a star around the
galaxy. Its motion will be determined by Eq.\ (\ref{lmo}) where
\[
F^{\perp}=G\frac{M_G m}{r^2}.
\]
For spiral galaxies  $F^\perp/m\ll a_0$ thus, since we are
in a MONDian regime, we have
\[
m (\frac{v^2}{r})^2 \frac{1}{a_0}=G\frac{M_G m}{r^2},
\]
which implies at once
\begin{equation}
v^4= \frac{G}{a_0} M_G.
\end{equation}
which is the Tully-Fisher relation. We recall that, more generally,
the Tully–-Fisher relation states $L \propto v^p$ where $L$ is the
luminosity of the galaxy. Observationally the wave-band dependent
exponent $p$ stays in the range [2.5, 5], and has the smallest
scatter in the near infrared for which $p$ is found to be close to
4, see \cite{mcgaugh00}. Observations give
\[
q a_p=a_0= 1.2 \times 10^{-10} m/s^2
\]
from which we obtain $q^{-1}\simeq 7$, thus $q$ is of the order of
unity as the consistency of our model required.

We conclude  that the perpendicular component of the friction force
can be chosen so as to reproduce all the important features of MOND.
Thus the many predictions of this theory can find a place in our
model. All that with the compatibility with Special Relativity in
the limit $a_p\to 0$.

\section{Conclusions}

We introduced an Aristotelian theory in which mass decreases
proportionally to the work done by the universal friction force on
the body. We showed that mass depends on velocity as in Special Relativity and that there is a characteristic velocity $c$ which is
insensitive to friction. Bodies with velocity smaller than $c$
decelerate till they reach a status of absolute rest. Bodies with
velocity $c$ preserve their velocity while they lose mass. We argued
that in the limit of vanishing friction $a_p\to 0$ the theory
becomes coincident with Special Relativity as the underlying
absolute space becomes unobservable. This result shows that Special Relativity, often regarded as incompatible with the concept of absolute (or ether) frame, can actually be obtained from a more detailed study of the interaction of bodies with such a frame.

In the last sections we confronted the
theory with experiment, showing that it can account for some
puzzling cosmological observations which involve accelerations of
the order of $10^{-10} \ m/s^2$. Although there are other phenomena that at present cannot be explained with this theory it seems worthwhile to investigate it further.





\begin{thebibliography}{10}

\bibitem{adler87}
C.~G. Adler.
\newblock Does mass really depend on velocity, dad?
\newblock {\em Am. J. Phys.}, 55:739--743, 1987.

\bibitem{Anderson:2001sg}
J.~D. Anderson, P.~A. Laing, E.~L. Lau, A.~S. Liu, M.~M. Nieto, and S.~G.
  Turyshev.
\newblock Study of the anomalous acceleration of {P}ioneer 10 and 11.
\newblock {\em Phys. Rev.}, D65:082004, 2002.

\bibitem{archard53}
J.F. Archard.
\newblock Contact and rubbing of flat surface.
\newblock {\em J. {A}ppl. {P}his.}, 24:981--988, 1953.

\bibitem{bach01}
C.~Bach.
\newblock {\em Die Maschinen-Elemente; ihre Berechung und Konstruktion}.
\newblock Bergstr{\"a}sser, Stuttgart, 1901.

\bibitem{baryshev01}
Yu.~V. Baryshev, A.~D. Chernin, and P.~Teerikorpi.
\newblock The cold local {H}ubble flow as a signature of dark energy.
\newblock {\em Astronomy and Astrophysics}, 378:729--734, 2001.

\bibitem{drabkin38}
I.~E. Drabkin.
\newblock Notes on the laws of motion in {A}ristotle.
\newblock {\em The American Journal of Philology}, 59:60--84, 1938.

\bibitem{dugas55}
R.~Dugas.
\newblock {\em A history of mechanics}.
\newblock Routledge and {K}egan {P}aul {LTD}., London, 1955.

\bibitem{ehlers73}
J.~Ehlers.
\newblock The nature and structure of space-time.
\newblock In J.~Mehra, editor, {\em The physicist's conception of nature},
  pages 51--91, Dordrecht, 1973. Raidel.

\bibitem{ehlers65}
J.~Ehlers, W.~Rindler, and R.~Penrose.
\newblock Energy conservation as the basis of relativistic mechanics {II}.
\newblock {\em Am. J. Phys.}, 35:995--997, 1965.

\bibitem{leibundgut96}
B.~Leibundgut {et al.}
\newblock Time dilation in the light curve of the distant type {Ia} supernova
  {1995K}.
\newblock {\em Astrophys. J.}, 466:L21--L26, 1996.

\bibitem{famaey12}
B.~Famaey and S.~S. McGaugh.
\newblock Modified {N}ewtonian {D}ynamics ({MOND}): {O}bservational
  phenomenology and relativistic extensions.
\newblock {\em Living Reviews in Relativity}, 15(10), 2012.

\bibitem{funaioli73}
E.~Funaioli.
\newblock {\em Corso di meccanica applicata alle macchine Vol. I}.
\newblock P\`atron, Bologna, 3rd edition, 1973.

\bibitem{heller06}
M.~Heller.
\newblock Evolution of space-time structures.
\newblock {\em Concepts of Physics}, 3:119--133, 2006.

\bibitem{holm46}
R.~Holm.
\newblock {\em Electrical contacs}.
\newblock H. {G}erber {P}ub., Stockholm, 1946.

\bibitem{irschik04}
H.~Irschik and H.J. Holl.
\newblock Mechanics of variable-mass systems-{P}art 1: Balance of mass and
  linear momentum.
\newblock {\em Appl. Mech. Rev.}, 57:145--160, 2004.

\bibitem{jammer00}
M.~Jammer.
\newblock {\em Concepts of Mass in Contemporary Physics and Philosophy}.
\newblock Princeton {U}niversity {P}ress, Princeton, 2000.

\bibitem{krasinsky04}
G.A. Krasinsky and V.A. Brumberg.
\newblock Secular increase of astronomical unit from analysis of the major
  planet motion, and its interpretation.
\newblock {\em Celestial Mechanics and Dynamical Astronomy}, 90:267--288, 2004.

\bibitem{landau62}
L.~D. Landau and E.~M. Lifshitz.
\newblock {\em The Classical Theory of Fields}.
\newblock {Addison-Wesley} {P}ublishing {C}ompany, Reading, 1962.

\bibitem{levyleblond76}
J.~M. L\'evy-Leblond.
\newblock One more derivation of the {L}orentz transformation.
\newblock {\em Am. J. Phys.}, 44:271--277, 1976.

\bibitem{levyleblond79}
J.~M. L\'evy-Leblond.
\newblock Additivity, rapidity, relativity.
\newblock {\em Am. J. Phys.}, 47:1045--1049, 1979.

\bibitem{mcgaugh00}
S.~S. McGaugh, J.~M. Schombert, G.~D. Bothun, and W.~J.~G. {de Blok}.
\newblock The baryonic {T}ully-{F}isher relation.
\newblock {\em Astrophys. J.}, 533:L99--L102, 2000.

\bibitem{milgrom83a}
M.~Milgrom.
\newblock A modification of the {N}ewtonian dynamics as a possible alternative
  to the hidden mass hypothesis.
\newblock {\em Astrophys. J.}, 270:365--370, 1983.

\bibitem{okun09}
L.~B. Okun.
\newblock Mass versus relativistic and rest masses.
\newblock {\em Am. J. Phys.}, 77:430, 2009.

\bibitem{opatowski42}
I.~Opatowski.
\newblock A theory of brakes, an example of a theoretical study of wear.
\newblock {\em Journal of the Franklin Institute}, 234, 1942.

\bibitem{parkyn58}
D.~G. Parkyn.
\newblock The effect of friction on elliptic orbits.
\newblock {\em The Mathematical Gazette}, 42:96--98, 1958.

\bibitem{penrose68}
R.~Penrose.
\newblock Structure of space-time.
\newblock In C.M. DeWitt and J.A. Wheeler, editors, {\em Battelle Rencontres},
  pages 121--235. Benjamin, New York–-Amsterdam, 1968.

\bibitem{penrose05}
R.~Penrose.
\newblock {\em The road to reality: A complete guide to the laws of the
  Universe}.
\newblock A. A. Knopf, New York, 2005.

\bibitem{penrose65d}
R.~Penrose and W.~Rindler.
\newblock Energy conservation as the basis of relativistic mechanics.
\newblock {\em Am. J. Phys.}, 33:55--59, 1965.

\bibitem{reye60}
T.~Reye.
\newblock Zur theorie der zapfenreibung.
\newblock {\em J. Der Civilingenieur}, 4:235--255, 1860.

\bibitem{sanders02}
R.~H. Sanders and S.~S. McGaugh.
\newblock Modified {N}ewtonian {D}ynamics as an alternative to {D}ark {M}atter.
\newblock {\em Ann. {R}ev. {A}stron. {A}strophys.}, 40:263--317, 2002.

\bibitem{sandin91}
T.~R. Sandin.
\newblock In defense of relativistic mass.
\newblock {\em Am. J. Phys.}, 59:1032--1036, 1991.

\bibitem{trautman70}
A.~Trautman.
\newblock Fibre bundles associated with space-time.
\newblock {\em Rep. Math. Phys.}, 1:29--62, 1970.

\bibitem{turyshev11}
S.~G. Turyshev, V.~T. Toth, J.~Ellis, and C.~B. Markwardt.
\newblock Support for temporally varying behavior of the {P}ioneer anomaly from
  the extended {P}ioneer 10 and 11 {D}oppler data sets.
\newblock {\em Phys. Rev. Lett.}, 107:081103, 2011.

\bibitem{turyshev12}
S.~G. Turyshev, V.~T. Toth, G.~Kinsella, S.-C. Lee, S.-M. Lok, and J.~Ellis.
\newblock Support for the thermal origin of the {P}ioneer anomaly.
\newblock {\em Phys. Rev. Lett.}, 2012.

\bibitem{villaggio01}
P.~Villaggio.
\newblock Wear of an elastic block.
\newblock {\em Meccanica}, 36:243--–249, 2001.

\end{thebibliography}


\end{document}